%% file: main.tex
\documentclass[conference]{IEEEtran}
\pdfoutput=1
\IEEEoverridecommandlockouts
\usepackage{cite}
\usepackage{amsmath,amssymb,amsfonts}
\usepackage{algorithmic}
\usepackage{graphicx}
\usepackage{textcomp}
\usepackage[pscoord]{eso-pic}
\usepackage[a4paper, total={184mm,239mm}]{geometry}
\usepackage{xcolor}

\usepackage{hyperref}
\hypersetup{hypertex=true,
            colorlinks=true,
            linkcolor=black,
            anchorcolor=black,
            citecolor=black} 
\usepackage{amssymb}
\usepackage{mathtools, nccmath}
\DeclarePairedDelimiter{\nint}\lfloor\rceil

\usepackage{subfigure}

\def\BibTeX{{\rm B\kern-.05em{\sc i\kern-.025em b}\kern-.08em
    T\kern-.1667em\lower.7ex\hbox{E}\kern-.125emX}}

\begin{document}
\title{Hardware Acceleration of Fully Quantized BERT for Efficient Natural Language Processing\\
}

\author{\IEEEauthorblockN{Zejian Liu$^{1,2}$, Gang Li$^{1}$, Jian Cheng$^{1,2}$}
\IEEEauthorblockA{
{$^1$National Laboratory of Pattern Recognition, Institute of Automation, Chinese Academy of Sciences} \\
{$^2$School of Future Technology, University of Chinese Academy of Sciences}\\
{Beijing, China} \\
{liuzejian2018@ia.ac.cn, \{gang.li, jcheng\}@nlpr.ia.ac.cn}}
}

\maketitle

\input{0-Abstract}
\input{1-introduction}

\input{3-Algorithm}

\input{4-Hardware}
\input{5-Results_and_discussion}
\input{6-Conclusion}
\input{7-Acknowledgment}
\input{8-Reference}

\end{document}

%% file: 0-Abstract.tex
\begin{abstract}
BERT is the most recent Transformer-based model that achieves state-of-the-art performance in various NLP tasks. In this paper, we investigate the hardware acceleration of BERT on FPGA for edge computing. To tackle the issue of huge computational complexity and memory footprint, we propose to fully quantize the BERT (FQ-BERT), including weights, activations, softmax, layer normalization, and all the intermediate results. Experiments demonstrate that the FQ-BERT can achieve 7.94$\times$compression for weights with negligible performance loss. We then propose an accelerator tailored for the FQ-BERT and evaluate on Xilinx ZCU102 and ZCU111 FPGA. It can achieve a performance-per-watt of 3.18 fps/W, which is 28.91$\times$ and 12.72$\times$ over Intel(R) Core(TM) i7-8700 CPU and NVIDIA K80 GPU, respectively. 
\end{abstract}


%% file: 1-introduction.tex
\section{Introduction}
In recent years, transformer-based models \cite{NIPS2017_7181} have been ubiquitous in various natural language processing (NLP) applications, including machine translation\cite{MT}, question answering and document analysis\cite{van_Aken_2019}. Compared with traditional recurrent neural network (RNN) and long short-term memory (LSTM)\cite{1997Long}, transformer-based model uses multi-head self-attention mechanism to encode and decode the information of input sentence to achieve better representative ability, and the Bidirectional Encoder Representations from Transformers (BERT)\cite{devlin-etal-2019-bert} is the state-of-the-art transformer-based model.

The architecture of BERT is depicted in Figure~\ref{BERT}. Different from the original transformer, BERT only consists of stacked encoder layers as it focuses on learning language representations, which is similar to convolutional neural network for computer vision. The building block of self-attention is essentially matrix-matrix multiplication, which is suitable for massive parallel computing, therefore high-performance GPUs are preferred to accelerate the training and inference in the cloud. However, the vast computational complexity ($>$20GFLOPs) and memory requirement ($>$320MB floating-point parameters) of BERT poses a significant challenge for resource-constraint platforms, which hinders the applications on edge devices.

In this paper, we investigate the acceleration of BERT on resource-constraint FPGAs through algorithm/hardware co-design. Specifically, to reduce memory footprint, we adopt quantization to compress the original model. Different from Q8BERT\cite{Zafrir2019} and Q-BERT\cite{Shen2019a}, which only quantize part of parameters, we propose a fully quantized BERT (FQ-BERT). FQ-BERT represents weights, activations, scale factors, softmax, layer normalization, and all the other intermediate results with integer or fixed-point format. Then we present an accelerator tailored for the proposed FQ-BERT and evaluate on two separate FPGAs. Experiments demonstrate that the proposed accelerator outperforms CPU and GPU by a large margin in terms of performance-per-watt with less than 1\% accuracy degradation on SST-2 dataset. The main contributions of this paper can be summarized as follows:
\begin{itemize}
\item We propose an efficient FPGA-based BERT accelerator via algorithm/hardware co-design. On the algorithm side, we propose a hardware-friendly FQ-BERT that represents all parameters and intermediate results with integer or fixed-point data type. On the hardware side, to support different operations during inference, we present an accelerator with dot-product based PE and bit-level reconfigurable multiplier for 8/4-bit and 8/8-bit multiplication at run-time.
\begin{figure}[]
\centerline{\includegraphics[width=\columnwidth]{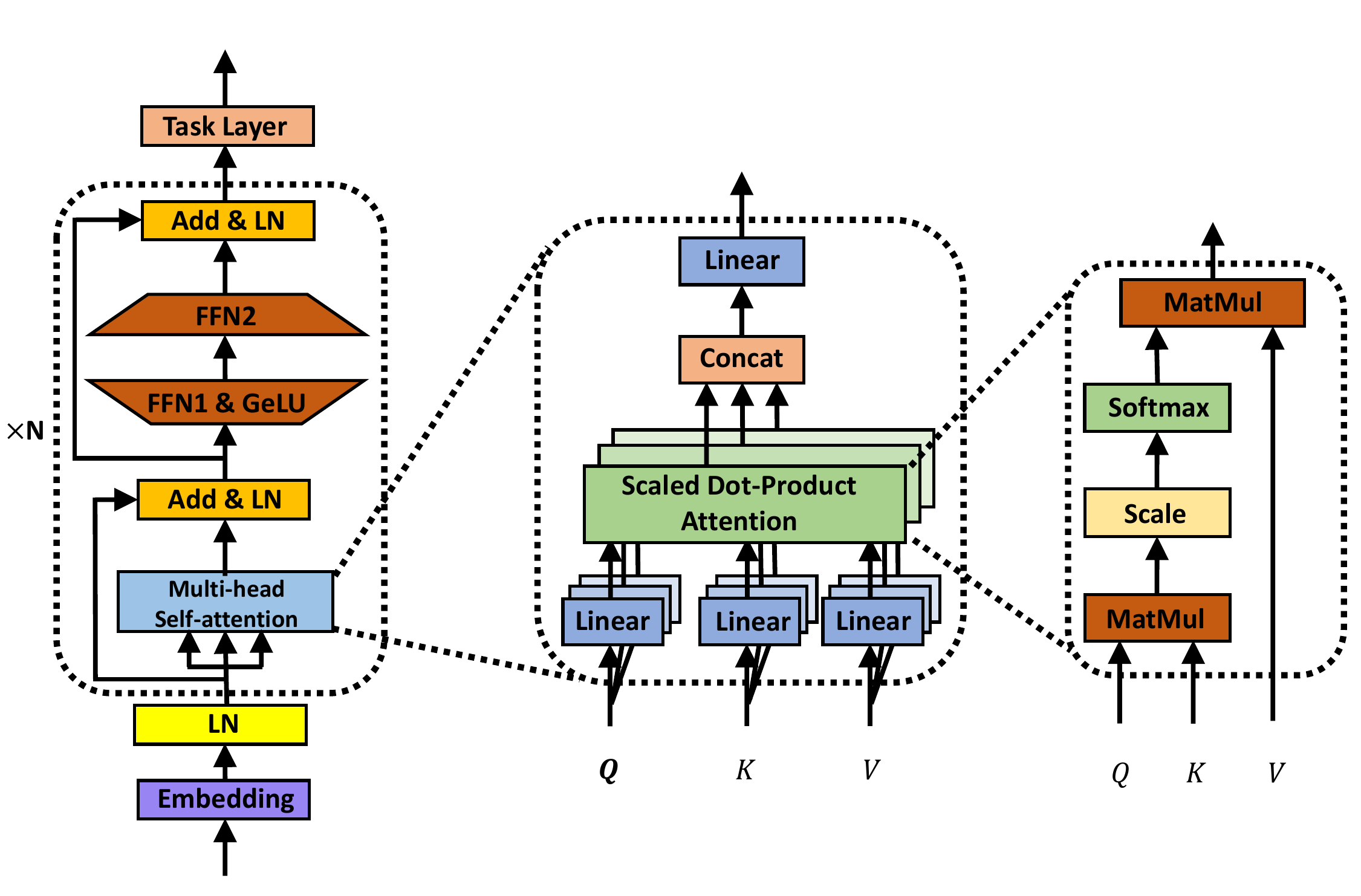}}
\caption{Left: the overall BERT architecture; middle: the architecture of multi-head self-attention module; right: the architecture of scaled dot-product attention layer.}
\label{BERT}
\end{figure}

\item We evaluate 8/4-bit FQ-BERT on SST-2 and MNLI dataset, which achieve a compression ratio of 7.94$\times$, with 0.7\% and 3.5\% accuracy drop respectively. Then we implement accelerator on Xilinx ZCU102 and ZCU111 MPSoC. Experiments demonstrate that our accelerator can achieve a performance-per-watt of 3.18 fps/W, which is 28.91$\times$ and 12.72$\times$ over Intel(R) Core(TM) i7-8700 CPU and NVIDIA K80 GPU, respectively. 
\end{itemize}

%% file: 3-Algorithm.tex
\begin{figure*}[htbp]
\centerline{\includegraphics{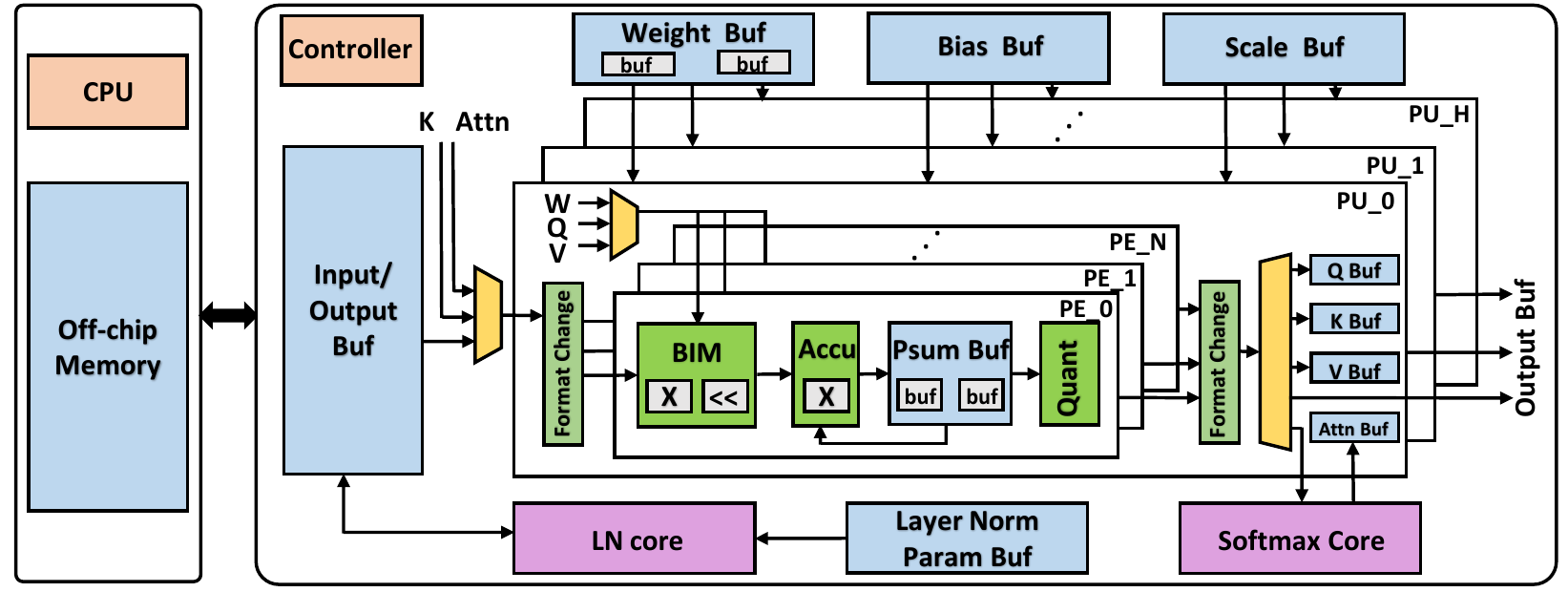}}
\caption{The overall architecture of the proposed accelerator for fully quantized BERT.}
\label{accelerator}
\end{figure*}

\section{Fully Quantized BERT}
In this section, we introduce the methods  and tricks that we used for quantizing parameters (weights, activations, biases and others) of the original BERT to obtain FQ-BERT.

\subsection{Weights and Activations}
\begin{figure}[]
    \centering
    
    \subfigure[SST-2]{
    \begin{minipage}{4.1cm}
    \centering
    \includegraphics[scale=0.25]{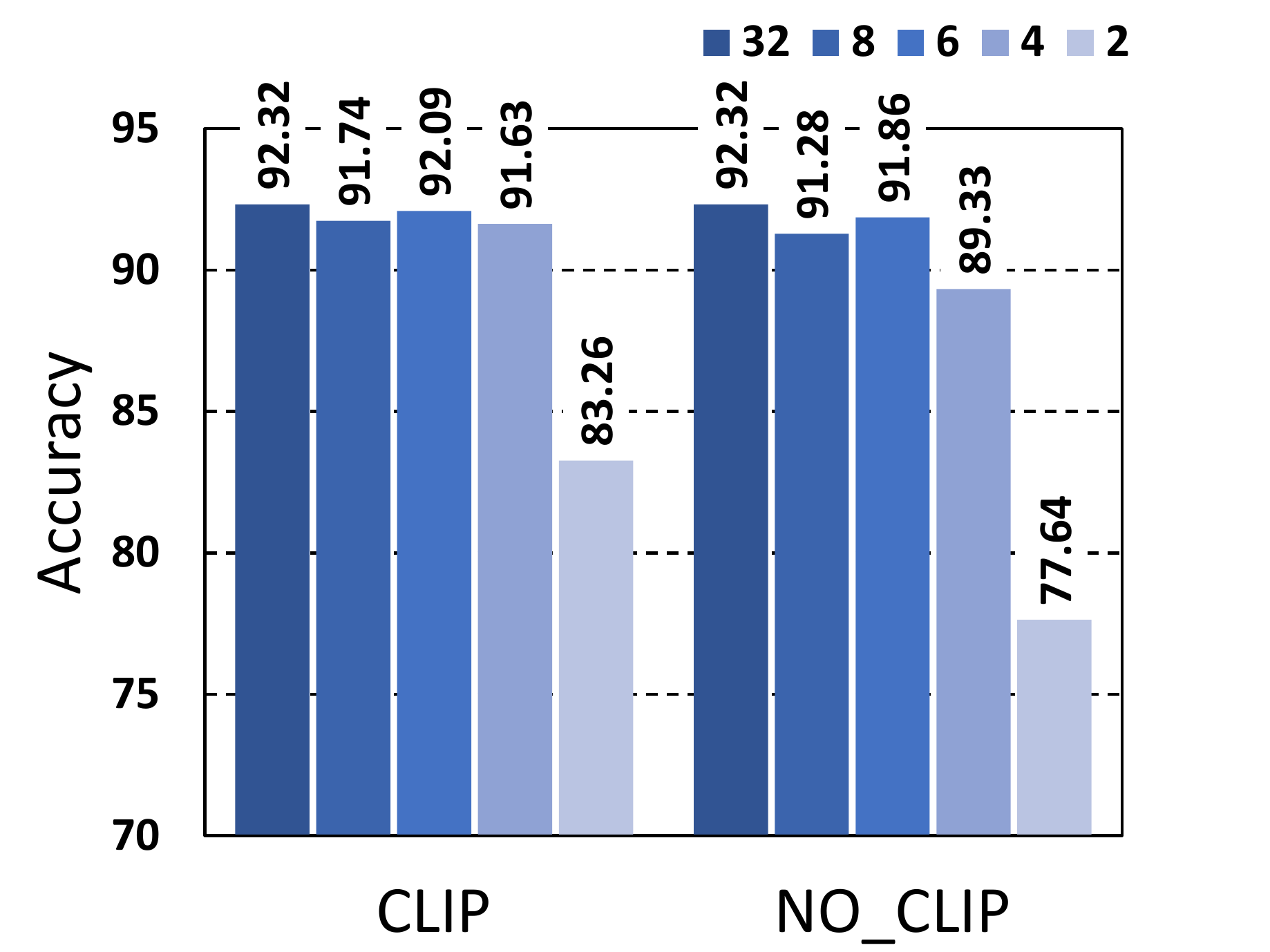}
    \end{minipage}
    }
    \subfigure[MNLI]{
    \begin{minipage}{4.1cm}
    \centering
    \includegraphics[scale=0.25]{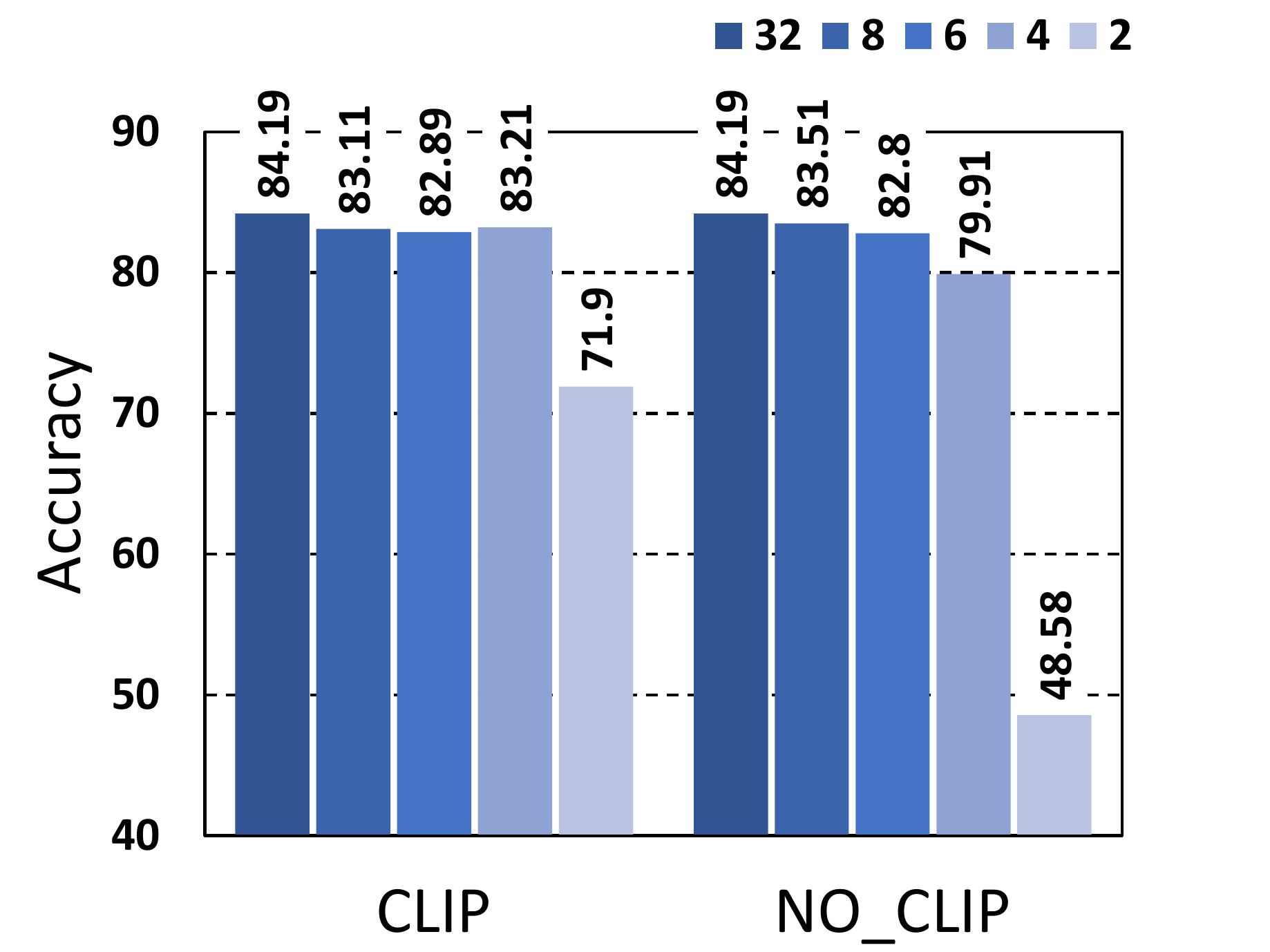}
    \end{minipage}
    }
    
    \caption{Impact of different quantization bitwidth of weights on accuracy.}
    \label{weight}
\end{figure}

To reduce the computational complexity and memory footprint, we adopt quantization to compress the BERT. In this work, we use symmetric linear quantization strategy for both weights and activations, because symmetric quantization is more hardware friendly for the lack of zero-point. Specifically, for each element $x$ of an input tensor, the $k$-bit quantization function is:

\begin{equation}
\begin{split}
    x_c &= clamp(x, MIN, MAX)  \\
    s &= scale(x_c, k)  \\
    x_I &= \nint{ x_c \times s} \\
    x_q &= x_I /s   \\
     clamp(x, a, b) &= min(max(x, a), b)
\end{split}
\end{equation}
Figure~\ref{weight} shows the quantization bitwidth of weights and corresponding accuracy on two document clssification datasets. It can be seen that when the bitwidth of weights is lower than 4, the classification accuracy drops dramatically, which indicates that 4-bit quantization for weights is a reasonable trade-off to maintain accuracy and considerable compression ratio. $MIN$ and $MAX$ are clip thresholds, which need to be carefully tuned during training. Its impact on performance is also depicted in Figure~\ref{weight}. 
Note that for symmetric quantization, $MIN=-MAX$. $s$ is a scaling factor, which is determined by the value of weights or activations and the quantization bitwidth $k$. For weights, the scaling factor is calculated according to:

\begin{equation}
\begin{split}
    s_w = scale(W, k) = \frac{2^{k-1}-1}{max(|W|)}
\end{split}
\end{equation}

We use exponential moving average (EMA) to collect the statistics of activations to determine scaling factor during inference:
\begin{equation}
\begin{split}
    s_a = scale(A, k) = \frac{2^{k-1}-1}{EMA(max(|A|))}
\end{split}
\end{equation}


\subsection{Biases and Others} 

Different from previous work that only weights and activations are quantized, our goal is to obtain a hardware-friendly BERT that can facilitate the deployment on hardware and improve the utilization of computing units. To this end, we further quantize all the biases to 32-bit integers with the help of weights' and activations' scaling factors:
\begin{equation}
\begin{split}
    bias_q &= \nint{bias \times s_{bias}} / s_{bias} \\
    s_{bias} &= s_a \times s_w
\end{split}
\end{equation}

As weights, activations and biases are all quantized, the calculation of the output can be rewritten as:
\begin{equation}
\begin{split}
    y_{I} =\nint{ y \times s_{y} }&= (\sum{a_I \times w_I} + b_I)\times s_f \\ 
    s_f& = \frac{s_y}{s_a \times s_w}
\end{split}
\end{equation}
where $s_a$, $s_w$ and $s_y$ are quantized to 8-bit values, and $s_f$ is a 32-bit integer. In addition, we also quantize the numerator and output of softmax module, and parameters of layer normalization to 8-bit fixed-point values.

%% file: 4-Hardware.tex
\begin{figure}[]
    \centering
    \subfigure[Type A]{
    \begin{minipage}{4.2cm}
    \centering
    \includegraphics[scale=0.21]{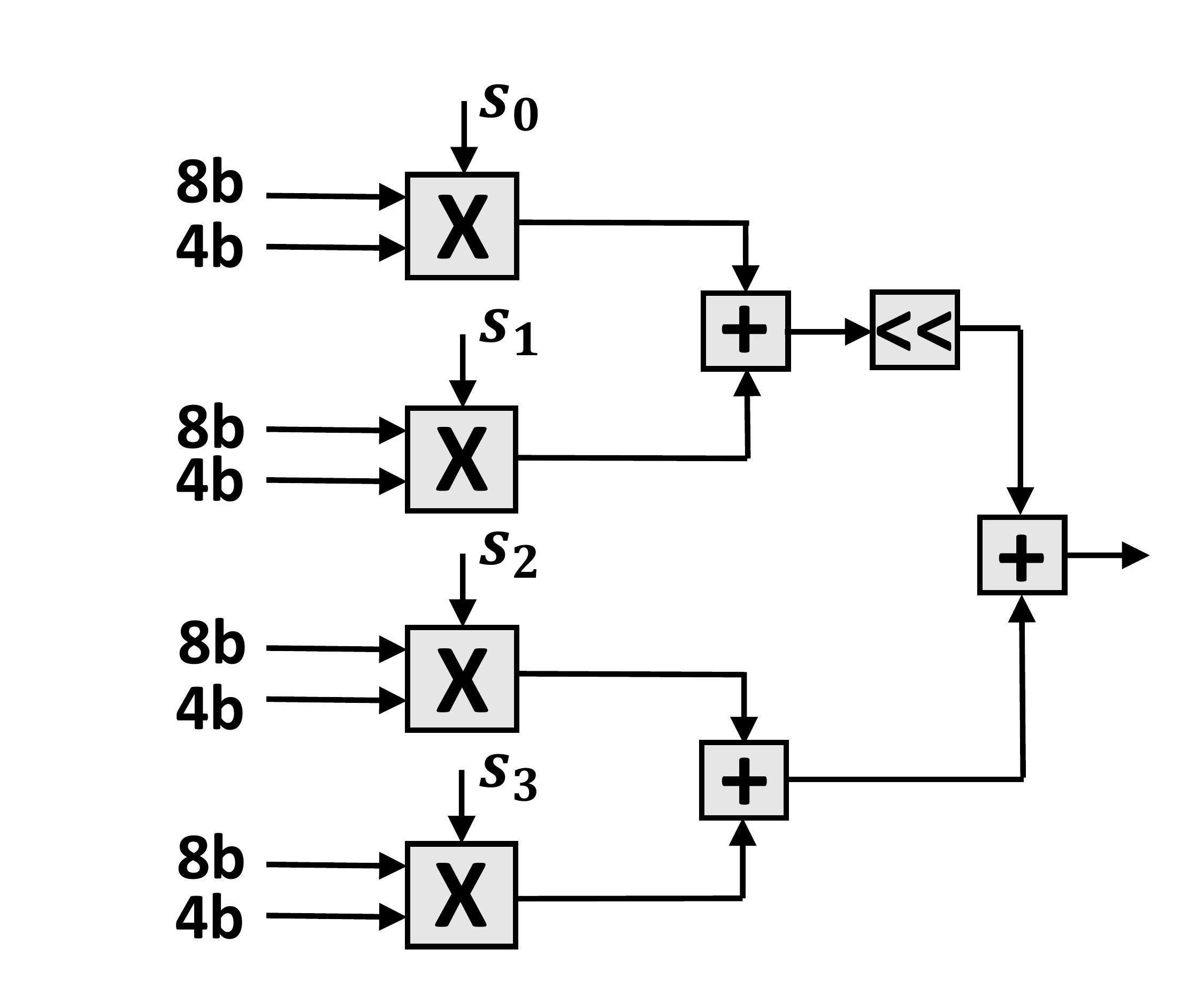}
    \end{minipage}
    }
    \subfigure[Type B]{
    \begin{minipage}{4cm}
    \centering
    \includegraphics[scale=0.21]{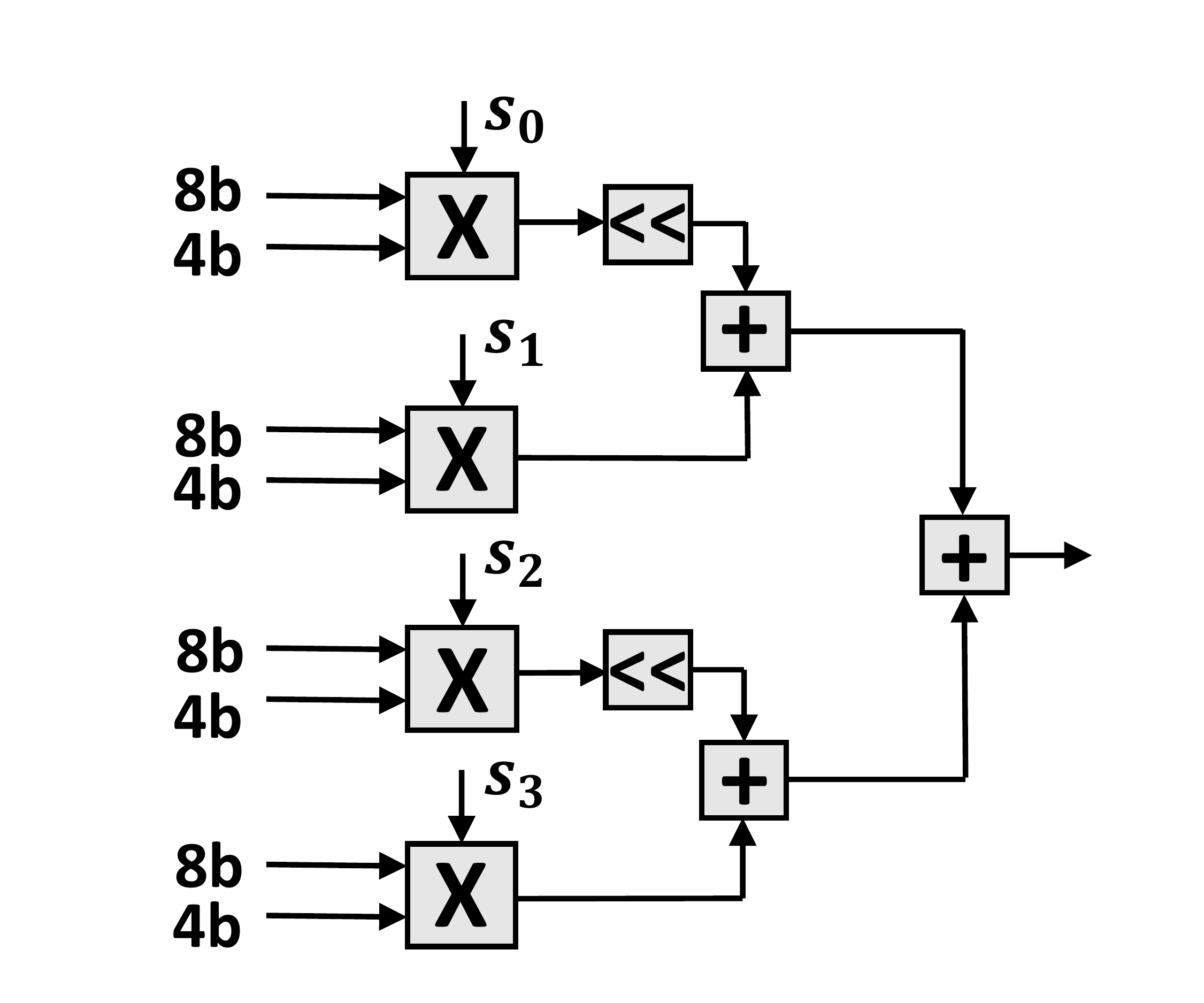}
    \end{minipage}
    }
    
    \caption{Two type of BIM. $M=4$. }
    \label{PE_MAC}
\end{figure}

\section{Accelerator}

\subsection{System Overview}
The overall architecture of our proposed accelerator is shown in Figure~\ref{accelerator}. It can be divided into two parts: the software program on  CPU and the hardware accelerator on FPGA.

As shown in Figure~\ref{BERT}, the whole model consists of embedding layers, encoder layers and task-specific layers. Different from encoder layers, the computation cost of other layers is small, but the memory cost is large, so these computations are completed on CPU, and the results are sent to FPGA. The weights are stored in off-chip memory and sent to FPGA through AXI4 interface. 

On the FPGA side, the accelerator consists of Processing Units (PUs), Layer Normalization (LN) Core, Softmax Core and On-chip Buffers. Each PU contains multiple Processing Elements (PEs), which is used for variable bitwidth matrix-vector multiplication. The On-chip Buffers include input/output buffer, weight buffer, parameter buffer and intermediate buffer. Specifically, the weight buffer is double buffered to overlap off-chip data transfer. The intermediate buffer is used for storing all the intermediate data, including $Q$, $K$, $V$ and attention matrix. The parameter buffer is used for caching data for special computation, such as scaling factors and softmax lookup table values. These data are loaded in the initialization procedure.

\subsection{Computing Unit }
\textbf{PE}. As different part of FQ-BERT has different bitwidth, the input bitwidth of multiplication has different combinations, such as 8-bit $\times$ 4-bit for $XW^Q$ and 8-bit $\times$ 8-bit for $QK^T$. In order to reuse the same module to support different operations, we design a Bit-split Inner-product Module (BIM), which is similar to Bit-fusion\cite{bitfusion}, as depicted in Figure~\ref{PE_MAC}.
Each BIM contains $M=2m$ 8-bit $\times$ 4-bit multipliers, two $m$-adder tree and several shift-add logic. Each multiplier has a sign signal to indicate the input number is signed or unsigned. The shift-add logic is used for shifting the partial sum when the bitwidth of input is larger than 4 bits. The position of shift-add logic has two choices, as shown in Fig.\ref{PE_MAC}, and using shift logic at adder tree's output (Type A) can save more resources, though this need to rearrange the input data. 

The output of BIM is a signed partial sum, which will be sent to an accumulator, as shown in Figure~\ref{accelerator}. The accumulator sums this input and previous data and send the results to the Psum Buf. When the calculation is completed, the results will be sent to quantization module to get final value incorporating with biases and scaling factors. Since the quantization process spends more than one cycle to complete, the Psum Buf is double buffered to ensure the calculation can be pipelined.

\textbf{Softmax Core}. 
BERT utilizes the softmax function to convert $QK^T$ to attention matrix, the expression of softmax is:
\begin{equation}
  Softmax(\boldsymbol{x})_{i} = \frac{exp(x_i)}{\sum_j{exp(x_j)}}
\end{equation}
whele $\boldsymbol{x}$ is a row vector of $QK^T$ matrix. As the exponential function $exp(x_i)$ is very expensive in terms of resource consumption, we use a lookup table instead. However, it is difficult to determine the number of entries of lookup table as the range of exponential function is very large and sensitive to input value. Fortunately, softmax is invariant to subtraction, i.e. subtract a constant for all the elements in the vector will not change the final results. If all the elements subtract the maximum value of the input vector, the output of exponential function will always be limited between 0 and 1. What's more, as we quantize $exp(x_i)$ to 8-bit, only 256 sampling points are needed.

\textbf{LN Core}. 
As most operations in layer normalization are element-wise multiplications and additions, with some non-linear operations, PE arrays are unsuitable for this computation pattern. We design a coarse-grained 3-stage pipelined SIMD unit to fully utilize its parallelism.  The first stage consumes two input vectors with two scaling factors and generates one vector and a mean value. The second stage subtracts the mean value from each element in the input and calculates variance value. The third stage finish the element-wise multiplication.

\begin{figure}[]
\centerline{\includegraphics[scale=0.26]{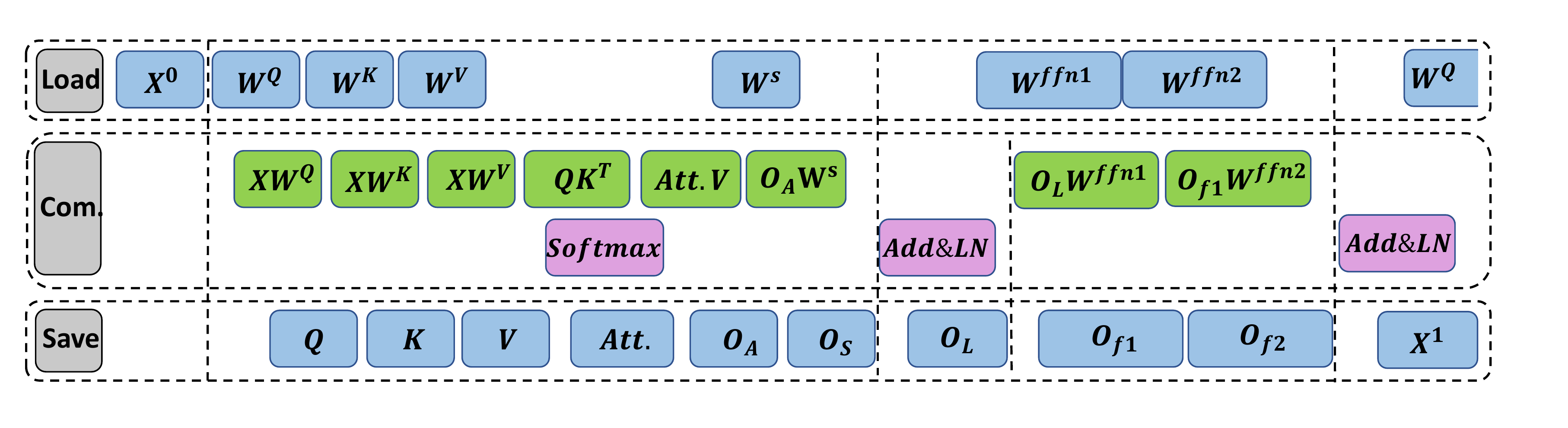}}
\caption{The dataflow of proposed accelerator.}
\label{dataflow}
\end{figure}

\subsection{Scheduling}
The dataflow of the accelerator is shown in Figure~\ref{dataflow}. According to different computation units and required weights, the computing process can be divide into several stages. Although the weights have been quantized to 4-bit, loading all the weights that needed per stage is still unpractical. So we further divide each stage into several sub-stages, and only load part of weights that will be used in next sub-stage. Through task-level scheduling, the off-chip transfer can be completely overlapped by computing.

%% file: 5-Results_and_discussion.tex
\section{Evaluation}
In this section, we evaluate the performance of proposed accelerator. First, we introduce the software and hardware environment setup of our experiments. Then, the performance of FQ-BERT is given. Finally, hardware resource utilization, latency and power consumption are provided. 

\subsection{Experimental Setup}
We implement the FQ-BERT using PyTorch and evaluate on two tasks, SST-2 and MNLI, of GLUE benchmark\cite{wang2018glue}. To get the quantized model, we first train the original model for 3 epochs with default hyper-parameters. Then we fine-tune the model with quantization function. 

The proposed hardware accelerator is built on Xilinx ZCU102 FPGA and ZCU111 MPSoC, and the frequency of FPGA part is set to 214 MHz. We implement the accelerator using High Level Synthesis. Vivado HLS 2019.1 is used for compiling the C++ code to generate synthesizable RTL. 

The baseline program runs on Intel(R) Core(TM) i7-8700 CPU @ 3.2GHz and NVIDIA K80 GPU with CUDA 10.1.  As we only consider the latency, all the experiments are running with batch size of 1, and the length of sentence is set to 128. 

\subsection{Accuracy}
The accuracy of FQ-BERT is shown in Table~\ref{acc}. For SST-2 task, the accuracy drop is 0.81\%, and for MNLI task, the accuracy drop is 3.08\% and 3.61\%, respectively. The accuracy loss on MNLI is larger than SST-2 as it is a more difficult task. After quantization, the model size is compressed 7.94$\times$. We also investigate the impact of different parts of BERT when they are quantized, the results are shown in Table~\ref{ablation}. We observe that the accuracy drop is not always increasing when more parameters are quantized, as quantizing softmax can improve the accuracy from 91.28\% to 91.86\%.

\begin{table}[]
\caption{Accuracy of FQ-BERT and baseline BERT.}
\begin{center}
\setlength{\tabcolsep}{2.5mm}{
\begin{tabular}{|c|c|c|c|c|c|}
\hline
 & \textbf{w/a} & \textbf{SST-2} & \textbf{MNLI} & \textbf{MNLI-m} & \textbf{Comp. Ratio} \\
\hline
\textbf{BERT} & 32/32 & 92.32 & 84.19 & 83.97 & 1$\times$\\
\hline
\textbf{FQ-BERT} & 4/8  &91.51 & 81.11 & 80.36& 7.94$\times$\\
\hline
\end{tabular}
}
\label{acc}
\end{center}
\end{table}

\begin{table}[]
    \centering
    \caption{Ablation study of quantization for different parts of BERT.}
    \setlength{\tabcolsep}{4mm}{
    \begin{tabular}{|c|c|c|c|c|}
        \hline
        \textbf{w/a} & \textbf{scale} & \textbf{softmax} & \textbf{layer norm} &  \textbf{accuracy}\\
        \hline
         -&- &- & -& 92.32\\
         \hline
         \checkmark &- &- & - & 91.63\\
         \hline
         \checkmark & \checkmark & - & - &91.28\\
         \hline
         \checkmark & \checkmark & \checkmark & - & 91.86\\
         \hline
         \checkmark & \checkmark & \checkmark & \checkmark & 91.51\\
         \hline
    \end{tabular}
    }
    \label{ablation}
\end{table}

\subsection{Resource Consumption}
Table~\ref{util} shows the resource consumption of the proposed accelerator for different configurations. Specifically, the number of PEs $N$ and the number of multipliers in BIM $M$ are examined. We can see that the DSP usage is very high for the targeted FPGA. To demonstrate the scalability of our accelerator, we double the number of total multipliers and implement it on ZCU111, and we get nearly twice the performance improvement.

\subsection{Performance and Energy Efficiency}
We further compare the performance and energy efficiency of our accelerator with CPU and GPU, results are shown in Table~\ref{speed}. Compared with CPU, our accelerator achieves 6.10$\times$  and 28.91$\times$ improvement in latency and energy efficiency. For GPU, our accelerator achieves 1.17$\times$ and 12.72$\times$ improvement. 

\begin{table}[]
\caption{Resource consumption and latency for different number of PEs and Multipliers in BIM. The number of PUs is 12.}
\begin{center}
\begin{tabular}{|c|c|c|c|c|c|}
\hline
\textbf{(N, M)} & \textbf{BRAM18K} & \textbf{DSP48E} & \textbf{FF} & \textbf{LUT} & \textbf{Latency(ms)}\\
\hline
\textbf{ZCU102} & 1824& 2520 &548160 & 274080 & -\\
\hline
\textbf{(8, 16)} & 838 & 1751& 124433 & 123157 & 43.89\\
\hline
\textbf{(16, 8)} & 877 & 1671 & 151010 &154192& 45.35\\
\hline
\hline
\textbf{ZCU111} & 2160 & 4272 & 850560 & 425280 & - \\
\hline
\textbf{(16, 16)} & 679$^{\mathrm{*}}$  & 3287 & 201469 &  189724 & 23.79\\
\hline
\multicolumn{6}{l}{$^{\mathrm{*}}$Some memory are implemented using URAM, which are not reported}
\end{tabular}
\label{util}
\end{center}
\end{table}

\begin{table}[]
\caption{Performance comparison on CPU, GPU and FPGA.}
\begin{center}
\setlength{\tabcolsep}{4mm}{
\begin{tabular}{|c|c|c|c|c|}
\hline
 & \textbf{CPU} & \textbf{GPU} & \textbf{ZCU102} & \textbf{ZCU111}\\
 \hline
\textbf{Latency(ms)}  & 145.06 & 27.84 & 43.89 & 23.79 \\
\hline
\textbf{Power(W)} & 65 & 143 & 9.8 & 13.2\\
\hline
\textbf{fps/W} & 0.11 & 0.25 & 2.32 & 3.18\\
\hline
\end{tabular}}
\label{speed}
\end{center}
\end{table}

%% file: 6-Conclusion.tex
\section{Conclusion}
In this paper, we investigate the acceleration of BERT on FPGAs. To reduce memory footprint, we propose to fully quantize all parameters of BERT, including weights, activations, scale factors, softmax, layer normalization and other intermediate results. Then we present an energy-efficient accelerator and evaluate it on two separate FPGAs. Experiments show that our accelerator outperforms CPU and GPU by factors of 28.91$\times$ and 12.72$\times$ on performance-per-watt respectively.

%% file: 7-Acknowledgment.tex
\section{Acknowledgment}
This work was supported in part by National Natural Science Foundation of China (No.61972396, 61876182, 61906193), National Key Research and Development Program of China (No. 2020AAA0103402), the Strategic Priority Research Program of Chinese Academy of Sciences (No. XDB32050200).

%% file: 8-Reference.tex
\bibliographystyle{ieeetr}
\bibliography{main}